\documentclass[12pt, letterpaper]{article}
\usepackage{arxiv}
\usepackage[utf8]{inputenc} 
\usepackage[T1]{fontenc}    
\usepackage{amsmath}
\usepackage[square,numbers]{natbib}
\usepackage{authblk}
\usepackage[pdftex,
pdfauthor={Mengjing Chen; Pingzhong Tang; Zihe Wang; Shenke Xiao; Xiwang Yang},
pdftitle={Optimal Anonymous Independent Reward Scheme Design}]{hyperref}

\title{Optimal Anonymous Independent Reward Scheme Design
}
\author[1]{Mengjing Chen\thanks{ccchmj@qq.com}\enspace}
\author[1]{Pingzhong Tang\thanks{kenshinping@gmail.com}\enspace}
\author[2,3]{Zihe Wang\thanks{wang.zihe@ruc.edu.cn}\enspace}
\author[1]{Shenke Xiao\thanks{xsk15@tsinghua.org.cn}\enspace}
\author[3]{Xiwang Yang\thanks{yangxiwang@bytedance.com}\enspace}
\affil[1]{Tsinghua University}
\affil[2]{Renmin University of China}
\affil[3]{Beijing Key Laboratory of Big Data Management and Analysis Methods}
\affil[4]{ByteDance}

\usepackage{microtype}
\usepackage{booktabs}
\usepackage{amsthm,amssymb}
\usepackage{bm}
\usepackage{physics}
\usepackage{tikz}
\usepackage{pgfplots}

\usepackage{algpseudocode}
\usepackage{algorithm}
\algnewcommand\algorithmicforeach{\textbf{for each}}
\algdef{S}[FOR]{ForEach}[1]{\algorithmicforeach\ #1\ \algorithmicdo}

\newtheorem{definition}{Definition}

\newtheorem{theorem}{Theorem}
\newtheorem{lemma}{Lemma}

\newcommand{\pre}{\mathrm{pre}}
\newcommand{\avg}{\mathrm{avg}}

\begin{document}

	\maketitle
\begin{abstract}
	We consider designing reward schemes that incentivize agents to create high-quality content (e.g., videos, images, text, ideas). The problem is at the center of a real-world application where the goal is to optimize the overall quality of generated content on user-generated content platforms. We focus on anonymous independent reward schemes (AIRS) that only take the quality of an agent's content as input. We prove the general problem is NP-hard. If the cost function is convex, we show the optimal AIRS can be formulated as a convex optimization problem and propose an efficient algorithm to solve it. Next, we explore the optimal linear reward scheme and prove it has a $\frac{1}{2}$-approximation ratio, and the ratio is tight. Lastly, we show the proportional scheme can be arbitrarily bad compared to AIRS. 
\end{abstract}

\section{Introduction}\label{intro}
User-generated content (UGC) platforms have become a major source for users to acquire information, knowledge, and entertainment. Representative platforms are video-sharing platform {\em YouTube}, question-and-answer platform {\em Quora}, online encyclopedia {\em Wikipedia} and lifestyle platforms {\em Instagram} and {\em TikTok}. According to several measurements~\cite{luca2015user,zhang2014differentiation,park2014exploring}, the UGC industry is just as important, if not more, as the search engine industry. 
These platforms are fundamentally different from search engines. The content returned by the former is generated by content providers (agents), while the latter returns mainly from professional authorities (at least for the first few pages). On the one hand, user-generated content presents diversity that is vital for the success of these platforms; on the other hand, the platforms face challenges in maintaining an overall high quality of the massive content generated. From an economic perspective, the platforms are required to design an incentive scheme that rewards high-quality content, given a restricted budget~\cite{ghosh2013learning,jain2014designing}. This is the central theoretical problem investigated in this paper.

The center designs a reward scheme, namely a {\em reward function} that maps a profile of agents' content (each with a real-valued quality score) to a reward profile. 
The goal is to maximize the overall quality of content on the platform with a fixed budget. 
The proportional scheme where each agent gets the reward is proportional to her contribution has been widely studied by many researchers~\cite{xia2014incentivizing,ghosh2014game,ghosh2011incentivizing,tullock1980efficient}. The proportional scheme has been proven easy to reach Nash equilibria among agents in the full information setting~\cite{xia2014incentivizing}.

In this paper,  we will focus on designing {\em anonymous independent reward schemes} (AIRS) in which the reward only depends on the quality of the individual content. Unlike the proportional mechanism, it does not depend on the quality of other agents' content either. The merit of the independent reward model is that it is easier for an agent to compute her best strategy. For the sake of fairness, we restrict the anonymity of the reward function, which is required by almost all real-world UGC platforms.

In our model, the agents have cost functions that are convex in their content quality to capture the idea that the additional cost to improve the marginal quality becomes heavier~\cite{ghosh2011incentivizing,ghosh2014game}.\footnote{The previous version considered the linear cost setting~\cite{chen2019optimal}.} Specifically, each agent has a type $t$ indicating the ability, and she costs $c(x)\cdot h(t)$ to produce content with quality $x$. When the type $t$ is higher, function $h(t)$ is lower. The type of users can be regarded as the skills of video clips or the design of the content. 
In this paper, we assume the quality of content entirely depends on the producer. 
We consider the Bayesian setting where each agent's ability is private information, and others only know its distribution.

In our model, the UGC platform has a limited budget to reward, such as money or web traffic. Given a simple reward function, an agent can choose how much effort to make to generate her content to maximize utility. Due to the incomplete information of each agent's ability, we assume the budget constraint is only required to be satisfied in expectation. The problem faced by the designer is then to optimize the sum of all content's quality by designing anonymous independent reward functions, subject to a fixed budget. Our model can be applied when a central organization needs to incentivize the community to accomplish some tasks. 

We model the problem of finding an optimal AIRS as an optimization problem in Section~\ref{Preliminaries}. In Section \ref{Convex Programming} we show that the optimization problem is equivalent to a convex optimization problem. Based on the analysis of the structure of the solution, we propose an efficient algorithm in Section~\ref{solution section}. 

In Section~\ref{sec: np-hard}, we claim the assumption of the cost function is somehow necessary since it is NP-hard to compute the optimal AIRS for general cost functions.
In Section~\ref{sec: linear}, we explore the linear reward scheme, which is a special case of the AIRS. We show the optimal linear reward scheme is a 2-approximation to the optimal AIRS and the ratio is tight. 

In Section~\ref{power} we exhibit the superiority of AIRS over other reward schemes. When agents are independent and identically distributed, the optimal AIRS beats any reward scheme implemented in a symmetric Bayes-Nash equilibrium. For the proportional reward scheme, we show it cannot be better than the optimal AIRS in the full information setting. It could be arbitrarily bad compared to the optimal AIRS.
\subsection{Related Work} \label{Related Works}
Our work contributes to the body of literature on pricing problem~\cite{o2005efficient,bjorndal2008equilibrium,azizan2019optimal}.  \citet{azizan2019optimal} consider the same setting but focus on dealing with agents' non-convex cost functions in an approximation way. We focus on the optimal solution in the convex cost setting. They allow different reward functions for different agents while we do not.

Our problem of designing optimal reward schemes is related to the principal-agent problem in contract theory. The principal designs compensation schemes (contracts) which incentivize agents to choose actions that maximize the principal's utility~\cite{holmstrom1982moral,babaioff2006combinatorial,dutting2021complexity}. The contracts they considered are different among agents while we design the common reward schemes. 
Nonetheless, there also exist researches on common contract design for multiple agents \cite{alon2020multiagent,xiao2020optimal}.
The significant differences between our model and the principal-agent model are: (1) the action spaces are usually finite and discrete in the principal-agent problem while we consider continuous action spaces, (2) we have a budget constraint in our model while principal-agent model does not have that.

Another line of literature is crowdsourcing, where individuals or organizations seek ideas or collect information from a large group of participants. The objective is to obtain ideas or solutions to some problems, and they only care about the best one ~\cite{chawla2019optimal,moldovanu2006contest}. In contrast, we are interested in the sum of the quality of the content. 

\section{Preliminaries} \label{Preliminaries}
Let $N=\{1,2,\dots, n\}$ be the set of all agents. Each agent $i$ has a type, denoted by a real number $t^i$, which stands for the ability of the agent to produce content. For each agent, the type is private information and drawn from a set $T_i$ with a probability mass function $f_i$. Set $T_i$ and function $f_i$ is publicly known. We assume $T_i$ is discrete and has a finite size. They are standard assumptions for compensation schemes design in the contract theory. We also define the union of type space $T=\bigcup_i T_i=\{t_1,\ldots,t_m\}$ and the sum of probabilities as $f(t)=\sum_{i=1}^n f_i(t)$. W.l.o.g., we assume $t_1<t_2<\cdots<t_m$.\footnote{Note that in this paper, the superscript refers to the agent while the subscript refers to the order of types.}
Every agent posts content (e.g., an article or a short video) on the platform. We use a non-negative number to represent the quality of the content (e.g., the expected number of views or likes) in which the platform is interested. We assume an agent can regulate the quality of her content and her action is choosing the quality. To produce content with quality $x\in \mathbb{R}_{\ge 0}$, an agent with type $t$ suffers from a cost $\mathrm{cost}(x,t)$. We consider cost functions in the form of  $\mathrm{cost}(x,t)=c(x)\cdot h(t)$ where $h(t)$ is always positive. We assume an agent with a higher type can produce content with the same quality using a lower cost, i.e., the function $h(t)$ is decreasing in $t$. We assume $c(x)$ is convex, strict increasing and $c(0)=0$. 

The platform designs a reward scheme, which is essentially a reward function $R:\mathbb{R}^n_{\ge 0}\mapsto \mathbb{R}^n_{\ge 0}$ that maps a quality profile of agents' content to a reward profile. We also define $R_i:\mathbb{R}^n_{\ge 0}\mapsto \mathbb{R}_{\ge 0}$ to be the reward function for agent $i$, i.e., $R(\bm{x})=(R_1(\bm{x}),\ldots,R_n(\bm{x}))$ where $\bm{x} = (x^1, x^2,\ldots, x^n)$. For agent $i$, we use $x^i$ $\in\mathbb{R}_{\ge 0}$ to represent her action of producing content with quality $x^i$ and use $\bm{x}^{-i}\in\mathbb{R}^{n-1}_{\ge 0}$ to represent other agents' actions similarly. 
An agent's utility is defined as the reward she receives minus the cost in producing her content, i.e.,  agent $i$'s utility function $u_i:\mathbb{R}^n_{\ge 0}\times T_i\mapsto \mathbb{R}$ that maps the quality profile of all agents' content and her type to her utility is defined as $u_i(x^i,\bm{x}^{-i},t^i)=R_i(x^i,\bm{x}^{-i})-c(x^i)h(t^i)$.

In this paper, we analyze the problem of incentivizing high-quality content in the {\em Bayesian information setting}. The platform aims to maximize the \emph{gross product} which is defined as the expectation of the overall quality of all content on the platform within a fixed budget $B$. We assume the budget constraint is only required to be satisfied in expectation. In addition, we assume the reward scheme satisfies individual rationality property so that an agent will not transfer money to the platform, i.e., the reward 
must be non-negative. 

We focus on a relatively simple reward scheme called {\it Anonymous Independent Reward Scheme} in which the reward each agent receives is only determined by the quality of content she produces and independent of other agents' actions. Any two agents receive the same reward if they create content with the same quality. The AIRS scheme has a simple format and can be easily understood by agents. Besides, it is anonymous thus has no price discrimination issue. 
\begin{definition}[Anonymous Independent Reward Scheme (AIRS)]
	A reward scheme is an Anonymous Independent Reward Scheme if 
	\begin{enumerate}
		\item the reward each agent receives is only determined by the quality of her content, i.e., for any $i$, $R_i(x^i,\bm{x}^{-i})=R_i(x^i,(\bm{x'})^{-i})$ for any $x^i\in\mathbb{R}_{\ge 0},\bm{x}^{-i},(\bm{x'})^{-i}\in \mathbb{R}_{\ge 0}^{n-1}$, and
		\item the reward function always assigns the same reward to any two agents if they produce content with the same quality, i.e., for any $i,j$, if $x^i=x^{j}$, $R_i(x^i,\bm{x}^{-i})=R_j(x^{j},\bm{x}^{-j})$ for any $\bm{x}^{-i}, \bm{x}^{-j}\in \mathbb{R}_{\ge 0}^{n-1}$.
	\end{enumerate}
	With a slight abuse of notation, we use function $R: \mathbb{R}_{\ge 0} \mapsto \mathbb{R}_{\ge 0}$ to represent the reward function in an AIRS. It only takes the quality of an agent's content as the input and specifies the agent's reward.
\end{definition}

We assume every agent is strategic and will produce content with the optimal quality to maximize her utility. Given the reward function, it is obvious that an agent's action only depends on her type. For convenience, we focus on an agent’s type instead of her index number from now on. We formulate the problem as an optimization problem presented as below. 

\begin{align}\label{P1}\tag{P1} 
\begin{aligned}
\max_{\substack{\mathbb{X}_k\subseteq \mathbb{R}_{\ge 0}\\G_k\in \Delta(\mathbb{X}_k)\\R}}  &&& \displaystyle \sum_{k\in [m]} f(t_k)\int_{x\in \mathbb{X}_k}x \dd{G_k(x)},\\
\text{s.t.} &&& 
R(x)-c(x)h(t_k)\ge R(y)-c(y)h(t_k), \\
&&&\qquad  \forall k\in [m], x\in \mathbb{X}_{k}, y\ge 0, \\
&&&  \displaystyle \sum_{k\in [m]}  f(t_k)\int_{x\in \mathbb{X}_k}R(x)\dd{G_k(x)}\le B,\\
&&&   R(x)\ge 0,\forall x\ge 0.
\end{aligned}
\end{align}
Here $\mathbb{X}_k$ represents the union of actions taken by all agents of type $t_k$, $\Delta(\mathbb{X}_k)$  represents the set of all cumulative probability distributions over $\mathbb{X}_k$, and $G_k$ is one cumulative probability distribution function over $\mathbb{X}_k$ that represents the combined mixed strategies used by agents of type $t_k$.  We will use a triple $(\mathbb{X}, G, R)$ to represent a solution to this problem.
The first constraint refers to agents choosing the best actions to maximize the utilities. Let $[m]$ represent the set $\{1,2,\ldots,m\}$ throughout the paper. The second constraint refers to the budget constraint. 
The last constraint refers to individual rationality. 

\section{The Optimal AIRS} \label{Convex Programming}
Problem \ref{P1} is complicated because it involves the design of mixed strategies $G_k$ and a reward function $R$ which involves a huge design space. To overcome these difficulties, we show that, w.l.o.g., we can assume the agents are using pure strategies. In addition, we reduce the design of the entire reward function to the design of the rewards on a set of specific values. At last, we show the optimal AIRS can be found by solving a convex optimization problem. 

We first state that an agent with a higher type will post content with (weakly) higher quality. 
\begin{lemma}\label{monotone_lemma}
	We pick two numbers $k,l \in [m]$ and assume $k<l$. Given a feasible solution $(\mathbb{X},G,R)$ to Problem \ref{P1}, for any $x_k\in \mathbb{X}_k$ and $x_l\in \mathbb{X}_l$, we have $x_k\le x_l$.
\end{lemma}
\begin{proof}
	By the best strategy constraints, we have $R(x_k)-c(x_k)h(t_k)
	\ge R(x_l)-c(x_l)h(t_k)$ and
	$R(x_l)-c(x_l)h(t_l)
	\ge R(x_k)-c(x_k)h(t_l)$.
	By summing up the two inequalities, we have $(x_k-x_l)(h(t_k)-h(t_l))\ge0$.
	Since $h(\cdot)$ is a decreasing function, we have $h(t_k)>h(t_l)$. Thus we get $x_k\le x_l$.
\end{proof}
Given a reward scheme, an agent might have multiple best actions and thus can use a mixed strategy. However, we can construct a solution where agents only use pure strategies. For any solution $(\mathbb{X},G,R)$ to Problem~\ref{P1}, we define $x_k=\int_{x\in\mathbb{X}_k}x \dd{G_k(x)}$ and $\Tilde{\mathbb{X}}_k=\{x_k\}$. We set 
$\Tilde{G}_k(x)=0$ for $x<x_k$ and $\Tilde{G}_k(x_k)=1$. In other words, an agent with type $t_k$ will produce content with quality $x_k$ deterministically. 
By Lemma~\ref{monotone_lemma}, $\mathbb{X}_k$
is point-wise weakly larger than $\mathbb{X}_{k-1}$.
Then the expectation of any distribution over $\mathbb{X}_k$ is weakly larger than the expectation of any distribution over $\mathbb{X}_{k-1}$, i.e., $x_k\ge x_{k-1}$.
Therefore it is proper to define $\Tilde{R}(x)=\max_y\{R(y)-c(y)h(t_k)\}+c(x_k)h(t_k)$ for $x\in[x_k, x_{k+1})$ where $x_{m+1}$ denotes infinity. We have the following result.

\begin{lemma}\label{lemma:tilde_r}
	Given a feasible solution $(\mathbb{X},G,R)$ to Problem \ref{P1}, $(\Tilde{\mathbb{X}},\Tilde{G},\Tilde{R})$ is also a feasible solution to Problem \ref{P1} and scheme $\Tilde{R}$ achieves the same gross product as scheme $R$. 
\end{lemma}
\begin{proof}
	We first show the solution $(\Tilde{\mathbb{X}},\Tilde{G},\Tilde{R})$ satisfies the first kind of  constraints in Problem \ref{P1}. It is obvious that an agent with any type will not choose action in $(x_k,x_{k+1})$ since the action $x_k$ is strictly better. 
	To show that $x_l$ is the best action of the agent with type $t_l$,
	it suffices to show the following for any $k\in[m]$:
	$$\Tilde{R}(x_k)-c(x_k)h(t_l)\le \Tilde{R}(x_l)-c(x_l)h(t_l).$$
	We prove the case $l>k$. The proof of the other case is similar and thus omitted.
	According to the definition of $x_k$, we can pick $x'\in \mathbb{X}_k$ such that $x'\ge x_k$.
	\begin{align*}
	&\Tilde{R}(x_l)-c(x_l)h(t_l)\\
	={}&\max_x\{R(x)-c(x)h(t_l)\}\\
	\ge{}& R(x')-c(x')h(t_l)\\
	={}&R(x')-c(x')h(t_k)+c(x')(h(t_k)-h(t_l))\\
	={}&\Tilde{R}(x_k)-c(x_k)h(t_k)+c(x')(h(t_k)-h(t_l))\\
	={}&\Tilde{R}(x_k)-c(x_k)h(t_l)+(c(x')-c(x_k))(h(t_k)-h(t_l))\\
	\ge{}&\Tilde{R}(x_k)-c(x_k)h(t_l).
	\end{align*}
	The third equality is for $x'\in \mathbb{X}_k$. The last inequality is for $h(t_k)\ge h(t_l)$.
	
	Next, we prove the budget constraint is still satisfied. By definition, $x_k=\int_{x\in \mathbb{X}_k}x \dd{G_k(x)}$. Since $c(\cdot)$ is convex, we have 
	$c(x_k)\le \int_{x\in \mathbb{X}_k}c(x) \dd{G_k(x)}$. According to the definition of $\Tilde{R}$, we have 
	$$\Tilde{R}(x_k)-c(x_k)h(t_k)=\int_{x\in \mathbb{X}_k} (R(x)-c(x)h(t_k))\dd{G_k(x)}.$$
	Combine them together, then we have
	$$\Tilde{R}(x_k)\le\int_{x\in \mathbb{X}_k} R(x)\dd{G_k(x)}.$$    
	
	At last, we compute the performance of scheme $\Tilde{R}$. 
	By the definition of $x_k$, we have $\sum_{k\in [m]}x_kf(t_k)=\sum_{k\in [m]}f(t_k)\int_{x\in \mathbb{X}_k}x \dd{G_k(x)}$. The proof completes.
\end{proof}

Function $\Tilde{R}$ is a step function. It is characterized by its break points $\{x_k\}$ and rewards on these points. It enables us to focus on these break points instead of the whole reward function. We continue to construct a new reward function $\hat{R}$ under which every agent's best action does not change, and the reward given by the platform weakly decreases. 
We define $\hat{R}(0)=0$ and $x_0=0$. For $x\in [x_k, x_{k+1}), k\in [m]$, we define
\begin{equation}
\hat{R}(x)=\hat{R}(x_{k-1})+(c(x_k)-c(x_{k-1}))h(t_k)  \label{hat R}.
\end{equation}
\begin{lemma}    \label{lemma hat R}
	Under reward function $\hat{R}$, $x_k$ is one best action for an agent with type $t_k$. 
	In addition, $\Tilde{R}(x_k)\ge \hat{R}(x_k)$ for  $k\in [m]$. 
\end{lemma}
\begin{proof}
	First, we verify that $x_k$ is the best action for an agent with type $t_k$.
	It suffices to prove for $k\neq l$
	$$\hat{R}(x_k)-c(x_k)h(t_k)\ge \hat{R}(x_l)-c(x_l)h(t_k)$$
	We only prove the case $k>l$, the proof of the case $k<l$ is similar.
	\begin{align*}
	&\hat{R}(x_k)-c(x_k)h(t_k)+c(x_l)h(t_k)\\
	={}&\hat{R}(x_{k-1})+(c(x_k)-c(x_{k-1}))h(t_k)-c(x_k)h(t_k)+c(x_l)h(t_k)\\
	={}&\hat{R}(x_{k-1})-c(x_{k-1})h(t_k)+c(x_l)h(t_k)\\
	={}&\hat{R}(x_{k-1})-c(x_{k-1})h(t_{k-1})-c(x_l)h(t_{k-1})+(c(x_l)-c(x_{k-1}))(h(t_k)-h(t_{k-1}))\\
	\ge{}& \hat{R}(x_{k-1})-c(x_{k-1})h(t_{k-1})-c(x_l)h(t_{k-1}).
	\end{align*}
	We repeat the above argument and finally we will get
	\begin{align*}
	\hat{R}(x_k)-c(x_k)h(t_k)+c(x_l)h(t_k)
	\ge{}&\hat{R}(x_{l})-c(x_{l})h(t_{l})-c(x_l)h(t_{l})\\
	={}&\hat{R}(x_l).
	\end{align*}
	
	Next, we prove $\hat{R}$ uses less reward compared to $\Tilde{R}$. 
	\begin{align*}
	&\Tilde{R}(x_l)-c(x_l)h(x_l)\ge \Tilde{R}(x_{l-1})-c(x_{l-1})h(x_l)\\
	\Rightarrow   {} &\Tilde{R}(x_l)-\Tilde{R}(x_{l-1})\ge (c(x_l)-c(x_{l-1}))h(x_l)\\
	\Rightarrow   {} &\Tilde{R}(x_l)-\Tilde{R}(x_{l-1})=\hat{R}(x_l)-\hat{R}(x_{l-1}).
	\end{align*}
	If we sum up the inequalities for $l\le k$ and using the fact $\Tilde{R}(x_1)>\hat{R}(x_1)$
	we will get $\Tilde{R}(x_k)\ge \hat{R}(x_k)$.
\end{proof}
\begin{figure}[!tbhp]
	\def\lasty{}
	\begin{center}
		\begin{tikzpicture}[
		thick,
		soldot/.style={only marks,mark=*},
		holdot/.style={fill=white,only marks,mark=*},
		every node/.style={transform shape}
		]
		\begin{axis}[xmin=0, xmax=9, ymin=0, ymax=15, 
		xtick={0,1,2,3,5.5,7},
		xticklabels={$0$,$x_1$,$x_2$,$x_3$,$x_{m-1}$, $x_m$},
		ytick={0},
		yticklabels={$0$},
		ylabel={$\hat{R}(x)$}, 
		xticklabel style={
			font=\Large
		},
		yticklabel style={
			font=\Large
		},
		ylabel style={
			font=\Large
		}
		]
		\foreach \xStart/\xEnd/\yVal  in 
		{0/1/0, 1/2/5, 2/3/7, 3/5.5/12, 5.5/7/13.5, 7/9/14.25
		} {
			\addplot[domain=\xStart:\xEnd, ultra thick] {\yVal}; 
			\ifx\lasty\empty\else
			\addplot[holdot](\xStart,\lasty);
			\addplot[soldot](\xStart,\yVal);
			\edef\tmp{\noexpand\draw[dotted]
				(axis cs: \xStart,\lasty) -- (axis cs: \xStart,\yVal);}
			\tmp
			\fi
			\global\let\lasty\yVal
		}
		\draw[dashed](0,0) -- (1,5) -- (2,7) -- (3,12) -- (5.5,13.5) -- (7,14.25);
		\end{axis}
		\end{tikzpicture}
	\end{center}
	\caption{Reward function $\hat{R}$ is a step function which has at most $m$ breakpoints.} \label{optimal reward function figure}
\end{figure}
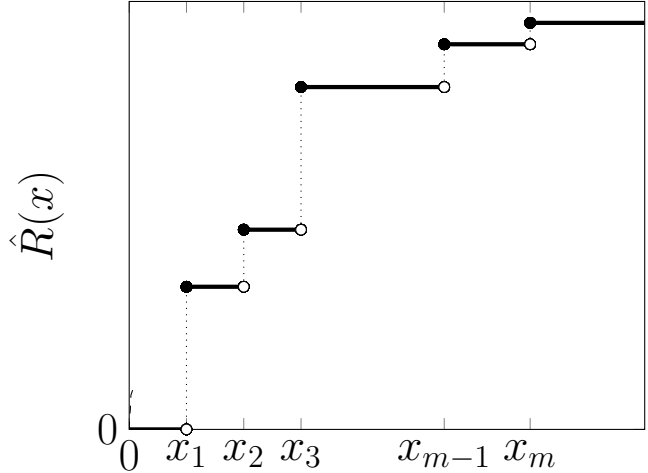
Given that $\Tilde{R}$ is budget feasible, the above lemma tells us scheme $\hat{R}$ is budget feasible. 
Furthermore, it suffices to find the optimal reward schemes $\hat{R}$ to solve Problem \ref{P1}. Figure \ref{optimal reward function figure} shows how function $\hat{R}$ looks like. Function $\hat{R}$ is fully characterized by the best action profile $\{x_k\}$. To simplify the problem, we rewrite $\hat{R}$ and constraints in terms of functions $c$, $h$ and $x_k$. By repeatedly using the definition of $\hat{R}$ in Eq. \eqref{hat R}, we will get 
\begin{align}
\hat{R}(x_k)=c(x_k)h(t_k)+\sum_{l=1}^{k-1}
c(x_{l})(-h(t_{l+1})+h(t_{l})).\label{hat_R}
\end{align}

The total reward given to agents can be written as  
\begin{align*}
&\sum_{k\in [m]}\hat{R}(x_k)f(t_k)\\
={}&\sum_{k\in [m]}f(t_k)\left(c(x_k)h(t_k)+\sum_{l=1}^{k-1}
c(x_{l})(-h(t_{l+1})+h(t_{l}))\right)\\
={}&\sum_{k\in [m]}c(x_k)\left(h(t_k)\sum_{l=k}^m f(t_l)-h(t_{k+1})\sum_{l=k+1}^m f(t_l)\right).
\end{align*}
In the second equality, we define $h(t_{m+1})=0$.
For the sake of simplicity, we define $\alpha_k$:
\begin{align}
\alpha_k= h(t_k)\sum_{l=k}^m f(t_l)-h(t_{k+1})\sum_{l=k+1}^m f(t_l).
\label{alpha k}
\end{align}
Since function $h(\cdot)$ is decreasing, we have $\alpha_k>0$.
At this point, we come to our first main result. 
\begin{theorem}
	\ref{P1} has the same optimal value as the following problem where $x_0=0$ and $\alpha_k$ is given in Eq. \eqref{alpha k}.
	\begin{align}
	\begin{aligned}
	\max_{x_1,\ldots,x_m} &&&\sum_{k\in [m]}x_kf(t_k),\\
	\mathrm{s.t.} &&& \sum_{k\in [m]}c(x_k)\alpha_k\le B,\\
	&&& x_{k-1}\le x_{k}, \forall k\in[m].
	\end{aligned}   
	\label{P2} \tag{P2}
	\end{align}
	Given \ref{P2}'s optimal solution $\{x^*_k\}$, we can construct the reward scheme $\hat{R}$ using Eq.~\eqref{hat R}.
\end{theorem}

Problem~\ref{P2} is a convex optimization problem. The objective is a linear function, and the feasible region is convex. Note that every point on the line segment connecting two feasible solutions is still in the feasible region since $c(x)$ is convex. 

\section{Solution Characterization and Algorithm}
\label{solution section}
In this section, we characterize the optimal solution of Problem \ref{P2} by analyzing the KKT conditions and thereafter propose an efficient algorithm to solve it. 
\subsection{Solution Characterization}
First note that the objective function is continuous and the feasible region is bounded and closed. Therefore, a global maximum exists.
We define the Lagrangian 
$$
\sum_{k\in [m]}x_kf(t_k)+\lambda\Big(B-\sum_{k\in [m]}\alpha_kc(x_k)\Big)+\sum_{k\in[m]}\mu_k(x_k-x_{k-1}).
$$
The KKT conditions for Problem~\ref{P2} are:
\begin{align}
f(t_k)-\lambda\alpha_kc'(x_k)+\mu_k-\mu_{k+1} &=0, k\in  [m],\label{kkt_0}\\
\lambda\Big(B-\sum_{k\in [m]}\alpha_kc(x_k)\Big)&=0, \label{kkt_2}\\
\mu_k(x_k-x_{k-1})&=0, k\in[m],\label{kkt_3}\\
\sum_{k\in[m]}\alpha_kc(x_k)&\le B,\label{kkt_5}\\
x_{k-1}-x_k&\le 0, k\in[m],\label{kkt_6}\\
\lambda\ge 0, \mu_k&\ge 0,k\in[m]. \label{kkt_7}
\end{align}
To handle the case where function $c(\cdot)$ does not have a derivative, we consider sub-derivatives such that $c'(x)$ can be any value in $[\partial_-c(x), \partial_+c(x)]$.

In Eq. \eqref{kkt_0}, we set $\mu_{m+1}=0$. If we sum up Eq.~\eqref{kkt_0} for all $k$, we get $\lambda>0$. In addition, by Eq. \eqref{kkt_2}, we deduce that there is no surplus in the budget.

To find the solution, it is important to determine the set of $k$s where $\mu_k=0$. We define $S=\{k\mid \mu_k=0, k\in [m+1]\}$. Note that $m+1\in S$. There exists a unique number, denoted as $q$, such that $\{1,\ldots,q\}\cap S=\{q\}$. 
For $k\in S$ and $k\neq q$, we define $\pre(k)$ as the ``predecessor'' element  such that $\{\pre(k), \pre(k)+1,\ldots,k-1\}\cap S=\{\pre(k)\}$. The ratio between the sum of $f$ and the sum of $\alpha$ shows a nice structure which can help us determine the set $S$. We first show a relation between the ratio and the dual variable $\lambda$. For convenience, we define 
$$\avg(l,k)=\frac{\sum_{j=l}^{k-1}f(t_j)}{\sum_{j=l}^{k-1}\alpha_j}.$$
\begin{lemma}
	For $1\le q<k$, $\avg(q,k)<\lambda c'(0)$.
	\label{bound on qs}
\end{lemma}
\begin{proof}
	By definition, we have $\mu_q>0$ for $q<k$. 
	According to Eq. \eqref{kkt_3} for $1\le q<k$, we have
	$x_0=x_1=x_2=\cdots=x_{q-1}$. Recall that $x_0$ is set to zero, therefore all these variables are zeros. 
	
	For any $k<q$, we sum up Eq. \eqref{kkt_0} from $k$ to $q-1$ and get
	
	
	$$\sum_{l=k}^{q-1}f(t_l)-\lambda\sum_{l=k}^{q-1}\alpha_l c'(0)+\mu_k=0.$$
	Since $\mu_k$ is positive, we have 
	$\avg(k,q)<\lambda c'(0) 
	$.
\end{proof}

We introduce some notations before further investigation. For $k\in [m]$, we define $\overline{\avg}(k)=\max_{1\le l<k}\{ \avg(l,k) \}$. For $2\le k\le m+1$, we define $\gamma(k)=\max\left\{l: \avg(l,k)= \overline{\avg}(k)\right\}$. The following theorem states a strong connection between functions $\pre(\cdot)$ and $\gamma(\cdot)$.

\begin{theorem}
	For any $k\in S$ and $k>q$, we have 
	$\pre(k)=\gamma(k)$. For $\gamma(k)\le j\le k-1$, variable $x_j$ has the same value. Furthermore, $\lambda c'(x_j)=\overline{\avg}(k)$.
	\label{S property}
\end{theorem}
The proof is in the same style as the proof for Lemma~\ref{bound on qs} but much more involved. 
\begin{proof}
	By definition of $S$, we have $\mu_{\pre(k)+1}$, $\mu_{\pre(k)+2}$, \ldots, $\mu_{k-1}$ are positive. By Eq. \eqref{kkt_3}, we have 
	$x_{\pre(k)}=x_{\pre(k)+1}=\cdots=x_{k-1}$. 
	By summing  up Eq. \eqref{kkt_0} from $\pre(k)$ to $k-1$, variables $\mu_k$ all cancel out and we have  $\sum_{l=\pre(k)}^{k-1}f(t_l)-\lambda\sum_{l=\pre(k)}^{k-1}\alpha_l c'(x_{\pre(k)})=0$. It implies
	\begin{align}
	\avg(\pre(k),k)=\lambda c'(x_{\pre(k)}). \label{ratio1}
	\end{align}
	If we sum up Eq. \eqref{kkt_0} from $k'$ to $k-1$ where $\pre(k)<k'<k$, we will have $\sum_{l=k'}^{k-1}f(t_l)-\lambda\sum_{l=k'}^{k-1}\alpha_l c'(x_{\pre(k)})+\mu_{k'}=0$.
	It implies 
	\begin{equation}
	\avg(k',k)<\lambda c'(x_{\pre(k)}).\label{ratio2}    
	\end{equation}
	Based on Eq. \eqref{ratio1} and \eqref{ratio2}, we have $\gamma(k)\le \pre(k)$.

	For $k'<\pre(k)$, we consider the intersection of the set $\{k',k'+1,\ldots,\pre(k)-1\}$ and $S$, denoted as $\{s_1,\ldots,s_j\}$. Then we have $s_j=\pre(\pre(k))$.
	Next we partition the set $\{k',\ldots,k-1\}$ into consecutive intervals: $\{k',\ldots,s_1-1\}$, $\{s_1,\ldots,s_2-1\}$,$\{s_2,\ldots,s_3-1\}$ and so on.
	According to the Eq. \eqref{ratio2}, we have
	\[\avg(k',s_1)\le \lambda c'(x_{\pre(s_1)})\le  \lambda c'(x_{\pre(k)})
	\]
	According to Eq. \eqref{ratio1}, for $1\le i<j$, we have
	\[\avg(s_i,s_{i+1})= \lambda c'(x_{\pre(s_{i+1})})\le \lambda c'(x_{\pre(k)}). \]
	Combining them together and making use of the definition of $\avg$, we get 
	\begin{align*}
	\sum_{l=k'}^{k-1}f(t_l)={}&\sum_{l=k'}^{s_1-1}f(t_l)+\sum_{l=s_1}^{s_2-1}f(t_l)+\cdots+\sum_{l=\pre(k)}^{k-1}f(t_l)\\
	\le{}&\lambda c'(x_{\pre(k)})\times
	\left(\sum_{l=k'}^{s_1-1}\alpha_l+\sum_{l=s_1}^{s_2-1}\alpha_l+\cdots+\sum_{l=\pre(k)}^{k-1}\alpha_l\right).
	\end{align*}
	It implies that 
	\[\avg(k',k)=\frac{ \sum_{l=k'}^{k-1}f(t_l)}{ \sum_{l=k'}^{k-1}\alpha_l}\le \lambda c'(x_{\pre(k)}).\]
	It indicates that $\overline{\avg}(k)=\lambda c'(x_{\pre(k)})$ and $\gamma(k)=\pre(k)$. The proof completes.
\end{proof}

For any integer larger than 1, function $\gamma(\cdot)$ maps it to a smaller integer, so there exists an integer $d$ such that $\gamma^{(d)}(m+1)=\gamma(\gamma(...\gamma(m+1)))=1$. Thus it is convenient for us to define $S_B=\{\gamma^{(d)}(m+1),\gamma^{(d-1)}(m+1)
,\ldots,m+1\}$. Lemma \ref{S property} tells us that for any $k>q$ in $S$, we have $\pre(k)=\gamma(k)$. Therefore, the set $S$ consists of every element in $S_B$ that is no less than $q$.

The next lemma states the monotonicity between any two consecutive elements in $S_B$. We design an $O(m)$-algorithm (Algorithm~\ref{alg:compute_s}) based on the monotonicity to compute $S_B$.
\begin{lemma}
	For any $k_1,k_2\in S_B$ and $k_1<k_2$, we have 
	$\avg(\gamma(k_1),k_1)\le \avg(\gamma(k_2),k_2)$.
	\label{SB property}
\end{lemma}
\begin{proof}
	As long as $\gamma(\gamma(k))$ is well defined for $k\in S_B$, by definition of $\gamma(k)$, we have
	\begin{equation*}
	\avg(\gamma(\gamma(k)),k)
	\le \avg(\gamma(k),k).
	\end{equation*}
	It implies that 
	\[\avg(\gamma(\gamma(k)),\gamma(k))
	\le \avg(\gamma(k),k).
	\qedhere
	\]
\end{proof}

\begin{algorithm}[htbp!]
	\caption{An $O(m)$ algorithm to compute $S_B$}
	\label{alg:compute_s}
	\textbf{Input}: $f(t_i)$, $\alpha_i$, $\forall i \in [m]$
	\begin{algorithmic} 
		\State Initilize empty stack $S_B$, AG and WT
		\ForEach {$k \in \mathcal [1,m] $}
		\State $\displaystyle\avg \gets \frac{f(t_k)}{\alpha_k}$, $ \text{weight} \gets \alpha_k$
		\While{$S_B$ is not empty}
		\If {$\text{AG.top} > \avg$}
		\State $\displaystyle \avg \gets \frac{ \avg \times \text{weight} + \text{AG.top}\times\text{WT.top}}{ \text{weight} + \text{WT.top}}$ 
		\State $\text{weight} \gets \text{weight} +\text{WT}$
		\State Pop($S_B$), Pop(AG) and Pop(WT)
		\Else
		\State break
		\EndIf
		\EndWhile
		\State Push($S_B$, $k+1$), Push(AG, $\avg$), Push(WT, $\text{weight}$)
		\EndFor
		\State Push($S_B$, $1$)
		\State \textbf{return} $S_B$
	\end{algorithmic}
\end{algorithm}

\begin{theorem}
	Algorithm \ref{alg:compute_s} computes $S_B$ in $O(m)$ time.
\end{theorem}
We provide the intuition here. The input can be equivalently regarded as a sequence of ratio $\frac{f(t_k)}{\alpha_k}$ with weight $\alpha_k$. While Algorithm \ref{alg:compute_s} is searching a set of break points to ``iron'' the given sequence into a non-decreasing sequence consists of $
\avg(\gamma(k),k)$. Note that each element can be pushed into the stack at most once and be popped out at most once. So the amortized time for each element in the array is $O(1)$. Then the total running time is $O(m)$ as a result. 

Given $S_B$, we still need $q$ to determine $S$. The next lemma gives a way to determine $q$ based on the value of $\lambda c'(0)$. The optimal solution of Problem \ref{P2} must satisfy $x_m>0$ and thus $q\le m$. Then we have $\lambda c'(0)\le \lambda c'(x_m)=\overline{\avg}(\gamma(m+1))$ which is an upper bound for $\lambda c'(0)$. 

\begin{lemma}
	If $\lambda c'(0)\in (\overline{\avg}(\gamma(k)),\overline{\avg}(k)]$ for some $k\in S_B$ such that $\gamma(k)>1, k\le m$, we have $q=\gamma(k)$.
	If $\lambda c'(0)\in [0, \overline{\avg}(k)]$ for the $k\in S_B$ such that $\gamma(k)=1$, we have $q=\gamma(k)$.
	\label{qs and gamma}
\end{lemma}
\begin{proof}
	We only prove the first claim. The proof for the second claim is the same. We prove it by a process of elimination. 
	We first suppose $q\ge k$.
	By definition of function $\gamma$, we have $\overline{\avg}(k)=\avg(\gamma(k),k)$. Moreover, we have
	\begin{equation*}
	\avg(\gamma(k),k)\le \avg(\gamma(q),q)<\lambda c'(0)
	\le \overline{\avg}(k).
	\end{equation*}
	The first inequality is based on Lemma~\ref{SB property}. The strict inequality is based on Lemma \ref{bound on qs}. Thus we derive 
	$\overline{\avg}(k)<\overline{\avg}(k)$, a contradiction.

	We next suppose $q<\gamma(k)$. Define $k'$ such that $\gamma(k')=q$ and then we have $k'\le \gamma(k)$. Furthermore, we have
	\begin{equation*}
	\overline{\avg}(\gamma(k))\ge 
	\overline{\avg}(k')=\lambda c'(x_q)\ge \lambda c'(0).
	\end{equation*}
	The first inequality is based on Lemma \ref{SB property}. The equality is based on Theorem \ref{S property}. We still get a contradiction.
	
	To sum up, it must be the case that $q=\gamma(k)$.
\end{proof}
Up to now, we only make use of constraints \eqref{kkt_0},\eqref{kkt_3},\eqref{kkt_6}, and \eqref{kkt_7} in KKT conditions. Given a parameter $\lambda$, if we leave alone the constraints \eqref{kkt_2} and \eqref{kkt_5}, we can determine $S$ and further find $\{x_k\}$ satisfying \eqref{kkt_0},\eqref{kkt_3},\eqref{kkt_6}, and \eqref{kkt_7}. The following lemma shows the influence of $\lambda$ on the corresponding solution $\{x_k\}$.

\begin{lemma}
	When the dual variable $\lambda$ increases, any $\{x_k\}$ that satisfies the constraints \eqref{kkt_0},\eqref{kkt_3},\eqref{kkt_6} and \eqref{kkt_7} decreases. 
	\label{xk and lambda}
\end{lemma}
\begin{proof}
	With $\lambda$ increasing, $q$ weakly increases according to Lemma~\ref{qs and gamma}. For a specific index $j$, as long as $q\le j$, $c'(x_j)$ decreases according to Theorem \ref{S property}. Note that there could be multiple solutions for $x_j$ given a specific $\lambda$. However, any solution is decreasing in $\lambda$. 
	Once $q>j$ happens, $x_j$ stays at zero. To sum up, $x_j$ decreases as $\lambda$ increases.
\end{proof}

Lemma \ref{xk and lambda} indicates an approach to finding the true $\lambda$. Recall that we would spend all the budget in the optimal solution. For a guess of $\lambda$, we can compute the $\{x_k\}$ and compare the total cost $\sum_{k\in [m]}\alpha_kc(x_k)$ and the budget $B$. If the total cost matches the budget $B$, the true $\lambda$ is found (Figure \ref{lambda figure}). 

Suppose there is a surplus in the budget. If $c'(\cdot)$ is not differentiable at $x_k$, we can tune down $\lambda$ unilaterally. If $c''(x_k)=0$, we can increase $x_k$ to a large value with $\lambda$ fixed.  If $c''(x_k)>0$, we can tune down $\lambda$ and increase $x_k$ simultaneously. In all three cases, we can spend more money. When there is a deficit in the budget, we can use similar methods to spend less money. Finally, the total cost would match the budget $B$. The next lemma gives a lower bound and an upper bound of $\lambda$ such that we can compute any approximation of the true $\lambda$ by the bisection method. 

\begin{lemma}
	\label{lambda bounds}
	Assume $c(y_1)=\frac{B}{\sum_k \alpha_k}$ and $c(y_2)=\frac{B}{\alpha_m}$, then we have 
	$$\frac{\overline{\avg}(\gamma(m+1))}{c'(y_2)}\le \lambda \le
	\frac{\overline{\avg}(\gamma(m+1))}{c'(y_1)}.$$
\end{lemma}
\begin{proof}
	We have argued that the equality holds in Eq. \eqref{kkt_5}. 
	On the one side, since $B=\sum_k \alpha_kc(x_k)\le c(x_m)\sum_k \alpha_k$, we have $x_m\ge y_1$.
	On the other side, since $B>\alpha_m c(x_m)$, we have $x_m\le y_2$.
	
	By Lemma \ref{S property}, $\lambda c'(x_m)=\overline{\avg}(\gamma(m+1))$. Plug in the two bounds of $x_m$, we get the bounds of $\lambda$.
\end{proof}
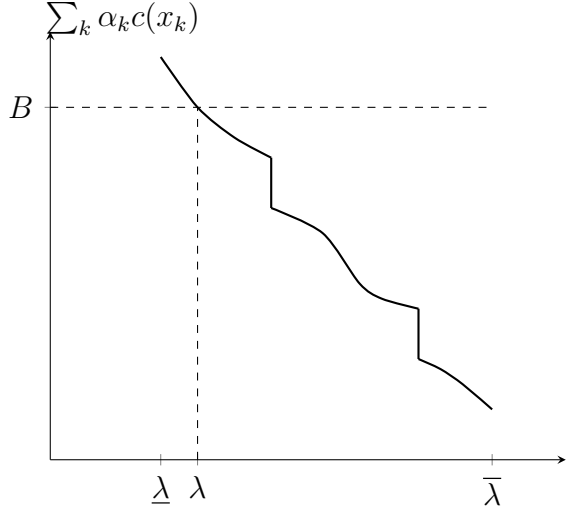
\begin{figure}
	\centering
	\begin{tikzpicture}
	\begin{axis}[
	axis lines = left, 
	xmin=0, xmax=7,
	ymin=0, ymax=8.5,
	ytick={7},
	yticklabels={$B$},
	xtick={1.5,2,6},
	xticklabels={$\underline{\lambda}$,$\lambda$,$\overline{\lambda}$},
	ylabel={$\sum_k \alpha_k c(x_k)$},
	y label style={
		at={(0.32,1.03)},
		rotate=-90,
	}
	]
	\addplot[mark=none, smooth, thick] coordinates{
		(1.5,8)
		(2,7)
		(2.5,6.4)
		(3,6)
	};
	\addplot[mark=none, thick] coordinates{
		(3,6)
		(3,5)
	};
	\addplot[mark=none, smooth, thick] coordinates{
		(3,5)
		(3.7,4.5)
		(4.2,3.5)
		(4.5,3.2)
		(5,3)
	};
	\addplot[mark=none, thick] coordinates{
		(5,3)
		(5,2)
	};
	\addplot[mark=none, smooth, thick] coordinates{
		(5,2)
		(5.3,1.8)
		(5.6,1.5)
		(6,1)
	};
	\addplot[mark=none, dashed] coordinates{
		(2,0)
		(2,7)
	};
	\addplot[mark=none, dashed] coordinates{
		(0,7)
		(6,7)
	};
	\end{axis}
	\end{tikzpicture}
	\caption{$\sum_k \alpha_k c(x_k)$ is decreasing in $\lambda$ on $[\underline{\lambda},\overline{\lambda}]$.} \label{lambda figure}
\end{figure}

We conclude this section by summarizing the algorithm to solve Problem \ref{P2}. 
\subsection{The Algorithm}
We first compute $S_B$ by Alg.~\ref{alg:compute_s}. Then we determine the lower bound and the upper bound of $\lambda$ by Lemma \ref{lambda bounds}. For a given $\lambda$, we can compute $q$ by Lemma \ref{qs and gamma} and the $\{x_k\}$ by  Theorem \ref{S property}. The true $\lambda$ satisfies the equation $\sum_k \alpha_k c(x_k)=B$. Since $\sum_k \alpha_k c(x_k)$ is decreasing in $\lambda$, the true $\lambda$ can be searched by bisection method. In the last step, there are two cases for the true $\lambda$. 
In one case, there is a unique solution profile $\{x_k\}$. 
In the other case, the possible values of 
$\sum_{k\in [m]}\alpha_kc(x_k)$ constitute a continuous interval due to the multiple solutions of $x_k$ given the value of $c'(x_k)$. Actually, for any solution satisfies the constraint $\sum_{k\in [m]}\alpha_kc(x_k)=B$ would be an optimal solution.

\begin{algorithm}[htbp!]
	\caption{Algorithm to solve \ref{P2}}
	\label{alg:compute_lambda}
	\textbf{Input}: $f(t_i)$, $\alpha_i$, $\forall i \in [m]$, $B$, $c(\cdot)$, $\epsilon$
	\begin{algorithmic} 
		\ForEach {$k \in \mathcal [m] $}
		\State $\overline{\avg}(k)=\max_{1\le l<k}\{ \avg(l,k) \}$
		\EndFor
		\State Compute $S_B$ by Algorithm \ref{alg:compute_s}
		\State $\gamma(m+1) = \max \{S_B/\{m+1\}\}$
		\State $\lambda_l = \underline{\lambda} = \gamma(m+1)/c'(c^{-1}(\frac{B}{\alpha_m}))$,
		\State $\lambda_h = \overline{\lambda} = \gamma(m+1)/c'(c^{-1}(\frac{B}{\sum_{k\in [m]}\alpha_m}))$
		\While{$\lambda_h - \lambda_l > \epsilon( \overline{\lambda} - \underline{\lambda} )$}
		\State $\lambda = (\lambda_h + \lambda_l) / 2$
		\State $S = \{k| \lambda c'(0) \ge \overline{\avg}(k), k \in S_B \}$
		\If {$S \ne S_B$}
		\State $S = S \bigcup \max \{S_B / S\}$
		\EndIf
		\State $k_0 = 1$
		\ForEach {$k \in S $}
		\State $x_i = (c')^{-1}(\overline{\avg}(k)/\lambda)$, $k_0 < i \le k$
		\State $k_0 = k$
		\EndFor		
		\If{$\sum_{k\in [m]} \alpha_k c(x_k) = B$} break
		\ElsIf{$\sum_{k\in [m]} \alpha_k c(x_k) > B$} $\lambda_l = \lambda$
		\Else  ~$\lambda_h = \lambda$
		\EndIf
		\EndWhile
		\State \textbf{return} $\lambda$, $\{x_k\}$
	\end{algorithmic}
\end{algorithm}

Let $\underline{\lambda}$ and $\overline{\lambda}$ denote the lower bound and the upper bound of $\lambda$ respectively given by Lemma \ref{lambda bounds}. We give the time complexity of Algorithm 2 for certain precision in the following theorem.
\begin{theorem}\label{runtime_lambda}
	Suppose that $\lambda^*$ is the optimal solution of Problem \ref{P2}. Given $\forall \epsilon > 0$, the run time of Algorithm 2 to find $\lambda$ such that $|\lambda - \lambda^*| < \epsilon |\overline{\lambda}-\underline{\lambda}|$ is $O(m\log\frac{1}{\epsilon})$.
\end{theorem}

\section{NP-Hardness}
\label{sec: np-hard}
In the two previous sections, we consider the convex cost function and propose an efficient algorithm to solve Problem \ref{P1}. If we relax the specific form of $c(x,t)$ and the convexity, the problem becomes difficult to solve, even in the full information setting. 
We consider the following cost function.
\begin{equation*}
\mathrm{cost}(x,t) = \begin{cases}
0, & 0\le x< 1,\\
t, & 1\le x\le 1+t,\\
+\infty, & 1+t<x.
\end{cases}
\end{equation*}
We show the decision version of the reward design problem with $\mathrm{cost}(x,t)$ is an NP-hard problem.
Given every agent type $t^i, i\in[n]$ and the above cost function, is there an AIRS that can achieve gross product $V$ within budget $B$? When there are multiple best actions, we assume an agent will choose the highest quality. We call this decision problem the ``General Cost Problem" for convenience.

\begin{theorem}\label{nphard}
	The General Cost Problem is NP-hard.
\end{theorem}
\begin{proof}
	We reduce the NP-complete problem the Subset Sum Problem to the General Cost Problem.
	Given a Subset Sum Problem instance where the positive integers are $(w_1,w_2,\ldots,w_n)$ and the value sum is $W$.
	We construct the reward problem by setting
	$t^i=w_i$ for $i\in [n]$,  $B=W$, and $V=W+n$.
	
	Suppose there is a solution $E$ for Subset Sum problem, i.e.,
	$\sum_{i\in E}w_i=W$. We assume $E=\{e_1,\ldots,e_{|E|}\}$ where $e_i$ is sorted in increasing order. Then we design an AIRS 
	\begin{equation*}
	R(x) = \begin{cases}
	0, & x< e_1+1,\\
	e_k, & 1+e_k\le x< 1+e_{k+1},\\
	e_{|E|}, & 1+e_{|E|}\le x.
	\end{cases}
	\end{equation*}
	If $i\in E$, agent $i$ will choose $1+e_i$ and get reward $e_i$.
	If $i\notin E$, agent $i$ will choose $1$ and get reward $0$.
	It is obvious that scheme $R$ is a solution to the General Cost Problem.

	Suppose there is a scheme $R$ to General Cost Problem. W.l.o.g., we only need to consider the reward function that is weakly increasing.
	According to the cost function, agent $i$ has only two possible best actions:  $1$ and $1+w_i$. We denote $E$ as the set of agents who choose larger than $1$. Then the gross product would be $n+\sum_{i\in E}w_i$. Since the reward should cover agents' cost, the sum of reward is at least $\sum_{i\in E} w_i$.
	To satisfy the budget constraint, we must have $\sum_{i\in E}w_i=W$, which implies a solution to the Subset Sum Problem.
\end{proof}

\section{The Linear Reward Scheme}
\label{sec: linear}
This section focuses on a simpler scheme where the reward function is linear, denoted by $R(x)=px$. Here $p$ is the per unit price of contribution. We will show the optimal linear reward function can achieve at least $\frac{1}{2}$ gross product of that achieved by the optimal AIRS. 
\begin{theorem}\label{linear thm}
	Optimal linear reward scheme is a $\frac{1}{2}$-approximation to the optimal AIRS. The ratio is tight.
\end{theorem}
\begin{proof}
	In the optimal linear reward scheme, let $y$ be the agent's action and $y_k$ be the agent's best action when the type is $t_k$. Then the utility $yp-c(y)h(t_k)$ achieves the maximum when $y=y_k$ which implies $p=c'(y_k)h(t_k)$, or $y_k=0$ and $p\le c'(y_k)h(t_k)$. W.l.o.g., we assume funciton $c(\cdot)$ is differentiable. Actually, if function $c(\cdot)$ is not differentiable we will consider the sub-derivatives and the proof does not change. 
	For convenience, we assume an agent could take any best action in the linear reward setting. This assumption will only improve the performance of linear reward scheme. Thus there is an optimal price $p$ and action profile $\{y_k\}$ such that there is no surplus in the budget. Our objective is to maximize the gross product within the budget.
	Before we compare the optimal solution of \ref{P2} and the optimal gross product of the linear reward scheme, we introduce the following problem.
	\begin{align}
	\begin{aligned}
	\max &&& \sum_k z_kf(t_k), \\
	\mathrm{s.t.} &&& \sum_k c(z_k)h(t_k)f(t_k)\le B,\\
	&&& z_{k-1}\le z_k, k\in[m],\\
	&&& z_0=0.
	\end{aligned}
	\label{P3}  \tag{P3}  
	\end{align}
	Recall the definition of $\alpha_k=h(t_k)\sum_{l=k}^m f(t_l)-h(t_{k+1})\sum_{l=k+1}^m f(t_l)$, we have $\alpha_k>h(t_k)f(t_k)$. As a result, any feasible solution of  \ref{P2} would be a feasible solution of  \ref{P3}. We have the optimal solution of   \ref{P2} cannot be better than that of  \ref{P3}. Next we compare the optimal solution of \ref{P3} and the optimal gross product of the linear reward scheme.
	
	Suppose we have $\sum_k z_kf(t_k)> 2\sum_k y_kf(t_k)$, then
	\begin{align}
	&\sum_k c(z_k)f(t_k)h(t_k) \nonumber\\
	\ge {}& \sum_k [c(y_k)f(t_k)+c'(y_k)f(t_k)(z_k-y_k)]h(t_k) \label{linear ineq1}\\
	\ge {}& \sum_k c'(y_k)f(t_k)(z_k-y_k)h(t_k) \nonumber\\
	\ge{}& p\sum_k(z_k-y_k)f(t_k)\label{linear ineq2}\\
	>{}&p\sum_k y_kf(t_k).\nonumber 
	\end{align}
	Inequality \eqref{linear ineq1} is based on the convexity of the cost function.  Inequality \eqref{linear ineq2} is based on the fact that $p=c'(y_k)h(t_k)$, or $y_k=0$ and $p\le c'(y_k)h(t_k)$. 
	In the optimal linear reward scheme, we have the budget constraint satisfied by an equality, i.e., $p\sum y_k f(t_k)=B$. 
	As a result, we have $\sum c(z_k)h(t_k)f(t_k)>B$. It contradicts to the budget constraint of  \ref{P3}. Hence, the assumption is false and we have $\sum_k z_k f(t_k)\le 2\sum_k y_k f(t_k)$.
	
	To sum up, we have the optimal solution of  \ref{P2} is at least half good as the optimal gross product of the linear reward scheme. 
	
	We show the ratio is tight by providing an example.
	There is only 1 agent and she has a unique type $t$ and $h(t)=1$. The budget is $1$.
	We let 
	\begin{equation*}
	c(x) = \begin{cases}
	\epsilon x, & x\le 1,\\
	(1+\epsilon)x-1, & 1<x.
	\end{cases}
	\end{equation*}
	$c(x)=[(1+\epsilon)x-(1+\epsilon)]_+$.
	We design an AIRS $R$ such that
	
	\begin{equation*}
	R(x) = \begin{cases}
	0, & x<\frac{2}{1+\epsilon},\\
	1, & x\ge \frac{2}{1+\epsilon}.
	\end{cases}
	\end{equation*}
	Under scheme $R$, the agent will choose $x=\frac{2}{1+\epsilon}$. We consider the linear reward scheme. If the price $p$ is set larger than $1$, according to the budget constraint, the agent's action $x$ should be at most $1$. If the price $p$ is set at most $1$, the agent's action $x$ is still at most $1$ according to the cost function. When $p=1$, agent's best action would be $1$. Thus the ratio between the two schemes is $1/\frac{2+\epsilon}{1+\epsilon}$, which approaches to $\frac{1}{2}$ when $\epsilon$ moves towards to zero.
\end{proof}

\section{Superiority over Other Schemes}\label{power}
This section demonstrates that the optimal AIRS gains high gross product compared to other reward schemes. We first show that when agents' types are independent and identically distributed, the optimal AIRS has superiority over other anonymous schemes implemented in symmetric Bayes-Nash equilibrium. Then we prove that the proportional scheme, which divides the reward according to the proportion of quality, can perform arbitrarily badly in the worst case. At last, since the Bayes-Nash equilibrium is not known for the proportional reward scheme, we only consider the full information setting and show the optimal AIRS beats the proportional reward scheme.
\begin{theorem}\label{anonymous reward scheme}
	When agents are independent and identically distributed, for any anonymous reward scheme $R$ in which the Nash equilibrium is symmetric, there is an AIRS $R'$ that can achieve at least the same gross product as $R$.
\end{theorem}
This theorem indicates that AIRS might be the optimal anonymous reward scheme. 
\begin{proof}
	We set $R'(x)=E_{\bm{x}^{-i}}[R(x^i=x,\bm{x}^{-i})]$, the expectation is taken over agent's strategies in Nash equilibrium. 
	For the same action, every agent's utility under scheme $R'$ equals to the her expected utility under scheme $R$. Thus her best response does not change. In other words, the two schemes achieve the same gross product. Note that 
	$E_{x^i}[R(x^i)]=E_{x^i}[E_{\bm{x}^{-i}}[R(x^i,\bm{x}^{-i})]]=E_{\bm{x}}[R(\bm{x})]$, the scheme $R'$ consumes the same budget as scheme $R$.
	Hence, scheme $R'$ is budget feasible and achieves the same gross product as $R$.
\end{proof}
Next, we focus on the proportional scheme \cite{ghosh2014game}. Formally, the utility of agent $i$ in this scheme can be represented as
\[ u_i(x^i,\bm{x}^{-i},t^i)=\dfrac{x^iB}{\sum_{j=1}^nx^j}-c(x^i)h(t^i).\]
For completeness, let $u_i(x^i,\bm{x}^{-i})$ be $0$ if $x^i=0$ for all $i$.
This scheme has no guarantee on the gross product compared to the optimal AIRS, even in the full information setting. 

\begin{theorem} \label{prop theorem}
	There are two agents, for any $\epsilon>0$, there exists $h(\cdot)$,$c(\cdot)$ and $(t^1,t^2)$ such that the Nash equilibrium $(x^{\mathrm{prop},1}, x^{\mathrm{prop},2})$ of the proportional scheme and for actions $(x^{*,1}, x^{*,2})$ achieved in the optimal AIRS, we have
	\[ x^{\mathrm{prop},1}+  x^{\mathrm{prop},2}\le \epsilon(x^{*,1}+x^{*,2}).\]
\end{theorem}
Theorem \ref{prop theorem} is proved by constructing an instance that the proportional scheme can perform arbitrarily bad compared to the optimal AIRS. 
The idea is as follows. When there is an agent $A$ with high ability and an agent $B$ with low ability, agent $A$ can get a big enough reward with mediocre content since agent $B$ has a low ability. 
\begin{proof}
	We let $h(t)=\frac{1}{t}$ and $c(x)=x$. We first compute the Nash equilibrium in this setting.
	According to the definition, we should have 
	\begin{align*}
	&0=\frac{\partial u_i(x^{\mathrm{prop},i},x^{\mathrm{prop},-i},t^i)}{\partial x^{\mathrm{prop},i}}\\
	\Rightarrow& \frac{x^{\mathrm{prop},i}}{x^{\mathrm{prop},i}+x^{\mathrm{prop},-i}}B-\frac{x^{\mathrm{prop},i}}{t^i}=0.
	\end{align*}
	By routine calculations, we get
	$$x^{\mathrm{prop},1}=\frac{Bt^1 t^1t^2}{(t^1+t^2)(t^1+t^2)},\quad x^{\mathrm{prop},2}=\frac{Bt^1t^2t^2}{(t^1+t^2)(t^1+t^2)}.$$
	The gross product would be $\frac{B t^1t^2}{t^1+t^2}$.
	
	Then we consider the following AIRS. We assume $t^1<t^2$ and set $R(x)=\min\left\{\frac{x}{t^2},B\right\}$. 
	The agent 1 will choose $x^{*,1}=0$ and agent 2 will choose $x^{*,2}=Bt^2$.
	
	As long as $t^1<\epsilon t^2$, we have
	$\frac{Bt^1t^2}{t^1+t^2}<Bt^2\epsilon$ which implies  $x^{\mathrm{prop},1}+x^{\mathrm{prop},2}<\epsilon(x^{*,1}+x^{*,2})$.
\end{proof}

Though the proportional scheme is not an AIRS, the following theorem shows that there is an AIRS such that agents choose the same actions across two schemes and this AIRS demands less budget. 
\begin{theorem}\label{prop_thm2}
	Given agents' types, we can design an AIRS that achieves the same gross product as in the proportional scheme. In addition, every agent chooses the same action across two schemes.
\end{theorem}
\begin{proof}
	Suppose $\bm{x}^{\mathrm{prop}}$ is the Nash equilibrium of proportional reward scheme. Without loss of generality, we assume $x^{\mathrm{prop},i} \le  x^{\mathrm{prop},j}$ for each $i < j$. 
	
	In the proportional reward scheme, we have 
	\begin{align*}
	0&=\frac{\partial u_i(x^{\mathrm{prop},i},\bm{x}^{\mathrm{prop},-i},t^i)}{\partial x^{\mathrm{prop},i}}\\
	&=
	\frac{\sum_{j\neq i}x^{\mathrm{prop},j}}{(\sum_{j}x^{\mathrm{prop},j})^2}B-c'(x^{\mathrm{prop},i})h(t^i).
	\end{align*}
	
	It implies 
	\begin{align*}
	\frac{1}{\sum_{j}x^{\mathrm{prop},j}}B&>c'(x^{\mathrm{prop},i})h(t^i),
	\end{align*}
	thus,
	\begin{align*}
	\frac{x^{\mathrm{prop},i}-x^{\mathrm{prop},i-1}}{\sum_{j}x^{\mathrm{prop},j}}B&>c'(x^{\mathrm{prop},i})(x^{\mathrm{prop},i}-x^{\mathrm{prop},i-1})h(t^i)\\
	&\ge(c(x^{\mathrm{prop},i})-c(x^{\mathrm{prop},i-1}))h(t^i).
	\end{align*}
	Note that Eq. \eqref{hat_R} still holds when $x_{k-1} \le x_{k}$ for $k \in [m]$. Thus we can rewrite $\hat{R}$ according to Eq. \eqref{hat_R}, and get
	\begin{align*}
	\hat{R}(x^{\mathrm{prop},k})&=\sum_{i=1}^k(c(x^{\mathrm{prop},i})-c(x^{\mathrm{prop},i-1}))h(t^i)\\
	&<\sum_{i=1}^k\frac{x^{\mathrm{prop},i}-x^{\mathrm{prop},i-1}}{\sum_j x^{\mathrm{prop},j}}B\\
	&=\frac{x^{\mathrm{prop},i}}{\sum_j x^{\mathrm{prop},j}}B.
	\end{align*}
	Then we have $\sum_{k=1}^n\hat{R}(x^{\mathrm{prop},k})<B$. It means the scheme $\hat{R}$ uses less budget. 
	According to Lemma \ref{lemma hat R}, action $x^{\mathrm{prop},i}$ is a best action for agent $i$ under reward scheme $\hat{R}$. This proves the second claim. 
\end{proof}

\section{Conclusion}
We consider designing anonymous reward schemes for platforms to maximize the overall quality of all content. This paper introduces the anonymous independent reward scheme. We first show the intractability of the general problems. Then, when the cost function is convex, we propose an efficient algorithm. We also give a tight approximation ratio for the optimal linear reward scheme compared to the optimal AIRS. Finally, we show the superiority of AIRS over other anonymous schemes under several settings. Many open problems remain in this research direction. How to compute the Bayes-Nash equilibrium in the proportional reward scheme in the Bayesian information setting? Will the reward scheme benefit from using the rank information? What is the optimal anonymous reward scheme?

\bibliographystyle{unsrtnat} 
\bibliography{ref}

\begin{thebibliography}{20}
\providecommand{\natexlab}[1]{#1}
\providecommand{\url}[1]{\texttt{#1}}
\expandafter\ifx\csname urlstyle\endcsname\relax
  \providecommand{\doi}[1]{doi: #1}\else
  \providecommand{\doi}{doi: \begingroup \urlstyle{rm}\Url}\fi

\bibitem[Luca(2015)]{luca2015user}
Michael Luca.
\newblock User-generated content and social media.
\newblock In \emph{Handbook of media Economics}, volume~1, pages 563--592.
  Elsevier, 2015.

\bibitem[Zhang and Sarvary(2014)]{zhang2014differentiation}
Kaifu Zhang and Miklos Sarvary.
\newblock Differentiation with user-generated content.
\newblock \emph{Management Science}, 61\penalty0 (4):\penalty0 898--914, 2014.

\bibitem[Park et~al.(2014)Park, Jang, Jaimes, Chung, and
  Myaeng]{park2014exploring}
Jaimie~Yejean Park, Jiyeon Jang, Alejandro Jaimes, Chin-Wan Chung, and
  Sung-Hyon Myaeng.
\newblock Exploring the user-generated content (ugc) uploading behavior on
  youtube.
\newblock In \emph{Proceedings of the 23rd International Conference on World
  Wide Web}, pages 529--534, 2014.

\bibitem[Ghosh and Hummel(2013)]{ghosh2013learning}
Arpita Ghosh and Patrick Hummel.
\newblock Learning and incentives in user-generated content: Multi-armed
  bandits with endogenous arms.
\newblock In \emph{Proceedings of the 4th conference on Innovations in
  Theoretical Computer Science}, pages 233--246. ACM, 2013.

\bibitem[Jain et~al.(2014)Jain, Chen, and Parkes]{jain2014designing}
Shaili Jain, Yiling Chen, and David~C Parkes.
\newblock Designing incentives for online question-and-answer forums.
\newblock \emph{Games and Economic Behavior}, 86:\penalty0 458--474, 2014.

\bibitem[Xia et~al.(2014)Xia, Qin, Yu, and Liu]{xia2014incentivizing}
Yingce Xia, Tao Qin, Nenghai Yu, and Tie-Yan Liu.
\newblock Incentivizing high-quality content from heterogeneous users: On the
  existence of nash equilibrium.
\newblock \emph{Proceedings of the AAAI Conference on Artificial Intelligence},
  28\penalty0 (1), 2014.

\bibitem[Ghosh and Hummel(2014)]{ghosh2014game}
Arpita Ghosh and Patrick Hummel.
\newblock A game-theoretic analysis of rank-order mechanisms for user-generated
  content.
\newblock \emph{Journal of Economic Theory}, 154:\penalty0 349--374, 2014.

\bibitem[Ghosh and McAfee(2011)]{ghosh2011incentivizing}
Arpita Ghosh and Preston McAfee.
\newblock Incentivizing high-quality user-generated content.
\newblock In \emph{Proceedings of the 20th international conference on World
  wide web}, pages 137--146. ACM, 2011.

\bibitem[Tullock(1980)]{tullock1980efficient}
Gordon Tullock.
\newblock Efficient rent seeking.
\newblock In \emph{Toward a Theory of the Rent-seeking Society.}, pages
  97--112. College Stations, TX:Texas A \& M University Pres, 1980.

\bibitem[Chen et~al.(2019)Chen, Tang, Wang, Xiao, and Yang]{chen2019optimal}
Mengjing Chen, Pingzhong Tang, Zihe Wang, Shenke Xiao, and Xiwang Yang.
\newblock Optimal mechanisms with budget for user generated contents.
\newblock \emph{arXiv preprint arXiv:1907.04740}, 2019.

\bibitem[O'Neill et~al.(2005)O'Neill, Sotkiewicz, Hobbs, Rothkopf, and
  Stewart~Jr]{o2005efficient}
Richard~P O'Neill, Paul~M Sotkiewicz, Benjamin~F Hobbs, Michael~H Rothkopf, and
  William~R Stewart~Jr.
\newblock Efficient market-clearing prices in markets with nonconvexities.
\newblock \emph{European journal of operational research}, 164\penalty0
  (1):\penalty0 269--285, 2005.

\bibitem[Bj{\o}rndal and J{\"o}rnsten(2008)]{bjorndal2008equilibrium}
Mette Bj{\o}rndal and Kurt J{\"o}rnsten.
\newblock Equilibrium prices supported by dual price functions in markets with
  non-convexities.
\newblock \emph{European Journal of Operational Research}, 190\penalty0
  (3):\penalty0 768--789, 2008.

\bibitem[Azizan et~al.(2020)Azizan, Su, Dvijotham, and
  Wierman]{azizan2019optimal}
Navid Azizan, Yu~Su, Krishnamurthy Dvijotham, and Adam Wierman.
\newblock Optimal pricing in markets with nonconvex costs.
\newblock \emph{Operations Research}, 68\penalty0 (2):\penalty0 480--496, 2020.
\newblock \doi{10.1287/opre.2019.1900}.
\newblock URL \url{https://doi.org/10.1287/opre.2019.1900}.

\bibitem[Holmstrom(1982)]{holmstrom1982moral}
Bengt Holmstrom.
\newblock Moral hazard in teams.
\newblock \emph{The Bell Journal of Economics}, pages 324--340, 1982.

\bibitem[Babaioff et~al.(2006)Babaioff, Feldman, and
  Nisan]{babaioff2006combinatorial}
Moshe Babaioff, Michal Feldman, and Noam Nisan.
\newblock Combinatorial agency.
\newblock In \emph{Proceedings of the 7th ACM conference on Electronic
  commerce}, pages 18--28. ACM, 2006.

\bibitem[Dutting et~al.(2021)Dutting, Roughgarden, and
  Talgam-Cohen]{dutting2021complexity}
Paul Dutting, Tim Roughgarden, and Inbal Talgam-Cohen.
\newblock The complexity of contracts.
\newblock \emph{SIAM Journal on Computing}, 50\penalty0 (1):\penalty0 211--254,
  2021.

\bibitem[Alon et~al.(2020)Alon, Dobson, Procaccia, Talgam-Cohen, and
  Tucker-Foltz]{alon2020multiagent}
Tal Alon, Magdalen Dobson, Ariel~D Procaccia, Inbal Talgam-Cohen, and Jamie
  Tucker-Foltz.
\newblock Multiagent evaluation mechanisms.
\newblock \emph{Proceedings of the AAAI Conference on Artificial Intelligence},
  34\penalty0 (02):\penalty0 1774--1781, 2020.

\bibitem[Xiao et~al.(2020)Xiao, Wang, Chen, Tang, and Yang]{xiao2020optimal}
Shenke Xiao, Zihe Wang, Mengjing Chen, Pingzhong Tang, and Xiwang Yang.
\newblock Optimal common contract with heterogeneous agents.
\newblock \emph{Proceedings of the AAAI Conference on Artificial Intelligence},
  34\penalty0 (05):\penalty0 7309--7316, 2020.

\bibitem[Chawla et~al.(2019)Chawla, Hartline, and Sivan]{chawla2019optimal}
Shuchi Chawla, Jason~D Hartline, and Balasubramanian Sivan.
\newblock Optimal crowdsourcing contests.
\newblock \emph{Games and Economic Behavior}, 113:\penalty0 80--96, 2019.

\bibitem[Moldovanu and Sela(2006)]{moldovanu2006contest}
Benny Moldovanu and Aner Sela.
\newblock Contest architecture.
\newblock \emph{Journal of Economic Theory}, 126\penalty0 (1):\penalty0 70--96,
  2006.

\end{thebibliography}

\end{document}